\documentclass[a4paper, 12pt, openany, oneside]{article}
\usepackage[cp1251]{inputenc} 
\usepackage[english, russian]{babel} 
\usepackage{amsmath, latexsym,  amssymb, array, graphics, eucal,
amsfonts, amsthm, amsmath,  amsfonts, array, bm}

\makeatletter

\newcommand{\diag}{\rm diag\,}

\makeatother
{
\usepackage{indentfirst}

\begin{document}
\large
\renewcommand{\refname}{\begin{center} REFERENCES\end{center}}
\newcommand{\E}{\mc{E}}
 \renewcommand{\abstractname}{\,}
 \makeatother

\begin{center}
{\bf The Kramers problem for Holway---Shakhov equation}
\end{center}

\begin{center}
  \bf A. V. Latyshev\footnote{$avlatyshev@mail.ru$}
\end{center}\medskip

\begin{center}
{\it Faculty of Physics and Mathematics,\\ Moscow State Regional
University, 105005,\\ Moscow, Radio str., 10--A}
\end{center}\medskip

\begin{abstract}
The analytical solution of a problem on isothermal sliding
of rarefied gas along a flat firm surface (the Kramers problem)
for Holway---Shakhov equation is presented.


\end{abstract}

\begin{center}
\bf 1.  Introduction
\end{center}

One of the first known problems of the kinetic theory, for
which the exact solution is received, is the Kramers problem,
or the problem about an isothermal flow of gas with sliding.
For this problem some its methods of the exact solution are constructed
(see, for example, \cite{1}, \cite{2}, \cite{3}, \cite{4},
\cite{5}, \cite{6}).

Interest to problems of  gas flow with sliding is important because its
solution allows to calculate, in particular, boundary conditions for the
Navier---Stokes equation.

The kinetic Holway---Shakhon equation was already used for
solution boun\-dary problems of the kinetic theory (see, \cite{8} - \cite{10}).

In work \cite{eq} parametres of Holway---Shakhon equation were
expressed through Prandtl number, self-diffusion coefficient and
kinematic viscosity.

Let's notice, that the kinetic ellipsoidal statistical equation
of Holway \cite{Holway} was
is applied to the solution of the Smoluchovsky problem about temperature jump
in work \cite{11}. In works \cite {12} -- \cite {14} were studied
boundary problems of the kinetic theory, in particular, the Kramers
problem for binary gases.

In work \cite{15} the analytical solution  of the Kramers problem
has been received for the ellipsoidal statistical equation with
frequency of collisions of the molecules, proportional to the module
of molecules velocity.

The analytical solution of problems about isothermal and thermal
sliding with accomodation boundary conditions has been received
in work \cite{16}.

Later in works \cite{17} and \cite{18} the Kramers problem was
generalized on the case of quantum gases. So, in work \cite{17}
the case of Fermi gases was considered, and in work \cite{18}
the case of Bose gases was considered.

In work \cite{19} have been entered moment boundary conditions for
boundary problems for the rarefied gas. In \cite{19} the Kramers
problem was solved for 2-moment boundary conditions.

On example of Kramers problem in works \cite{20} -- \cite{23} were
effective methods of the approached solution of boundary problems
of kinetic theory are deve\-lo\-ped.

Let the half-space $x> 0$ is occupied by the one-nuclear
rarefied gas, a plane
$yz$ is combined with a wall, gas moves in the axis direction $y$ with
the mass velocity $u_y(x)$. Far from a wall
the constant of a gradient of mass velocity of gas is set
$$
g_v=\left(\dfrac{du_y(x)}{dx}\right)_{x=+\infty}.
\eqno{(1.1)}
$$

The given gradient of mass velocity of gas far from a wall causes so-called
sliding of gas along a wall with some unknown speed of sliding,
proportional quantity of the gradient of mass speed. In the Kramers
problem is required to define this unknown velocity of sliding, to construct
function of distribution of gas molecules and to find distribution mass
velocity of gas in half-space.

Let's notice, that the gradient of mass velocity $g_v$ and the gradient
of dimensionless mass velocity $U_y(x_1)=\sqrt{\beta}u_y(x_1)$
(on dimensionless coordinate $x_1 =\nu \sqrt{\beta}x$)
are connected by the relation
$$
G_v=\dfrac{g_v}{\nu}.
$$

Let's consider, that for the dimensionless gradient $G_v=g_v/\nu $
the inequality is carried out
$$
G_v\ll 1.
\eqno{(1.2)}
$$

The condition (1.2) allows to consider Kramers problem in the linear
statement.

From (1.1) follows, that for the mass
velocity in half-space $x>0$ far from boundary looks like
$$
u_y(x)=u_0+g_vx+o(1), \quad x\to +\infty,
\eqno{(1.3)}
$$
where the quantity $u_{sl}$ is called as velocity of isothermal
slidings along a flat surface.
Dimensionless velocity of sliding is equal $U_{sl}=\sqrt{\beta}\, u_{sl}$.

As the boundary condition on a wall we will accept the condition
purely diffusion reflexion of molecules from a wall
$$
f(x=0,y,z,\bm v)=f_0(v), \qquad v_x>0.
\eqno{(1.4)}
$$
Here $f_0(v)$ is the absolute Maxwellian,
$$
f_0(v)=n\Big(\dfrac{m}{2\pi kT}\Big)^{3/2}
\exp\Big[-\dfrac{mv^2} {2kT}\Big],
$$
i.e. the molecules reflected from a wall have Maxwellian
distribution by the velocities.

Concentration of gas and temperature in the Kramers problem
about isothermal sliding are considered as constants.

In the Kramers problem it is possible to consider distribution
function depen\-ding from velocity of molecules $\mathbf{v}$ and
one spatial coordinate $x $. Distribu\-tion function we will search in
the form
$$
f(x,\mathbf{v})=f_0(v)[1+h(x,\mathbf{C})].
\eqno{(1.5)}
$$

The mass velocity of gas directed along an axis $y $, looks like
$$
u_y(x)=\dfrac{1}{n}\int v_y\,f(x,\mathbf{v})\,d^3v.
\eqno{(1.6)}
$$

Boundary condition on the wall under the condition diffusion
reflexions of molecules
from the wall we will receive, if we will substitute decomposition (1.5) in
condition (1.4).  As result we receive the first boundary condition
(condition on the wall)
$$
h(0,C_x)=0, \qquad C_x>0.
\eqno{(1.7)}
$$

The second boundary condition (condition "far from the wall")
follows from this the fact,
that at $x\to + \infty $ distribution function
$h(x_1, \mathbf{C})$ passes into asymptotic function
$h_{as}(x_1, \mathbf{C})$.
Asymptotic function should be solution of  the Holway---Shakhov
equation.

\begin{center}
{\bf 2. The Kramers problem for Holway---Shakhov equation}
\end{center}

For Kramers problem the Holway---Shakhov equation becomes simpler and
has the following form \cite{eq}
$$
C_x\dfrac{\partial h}{\partial
x}+h(x,\mathbf{C})=$$$$=2C_yU_y(x)+\gamma
C_y\Big(C^2-\dfrac{5}{2}\Big)Q_y(x)+2\omega C_xC_yP_{xy}(x).
\eqno{(2.1)}
$$

In this equation the variable $x$ is the dimensionless coordinate,
connected with dimensional coordinate $x_1$ by the relation $x =\nu
\sqrt {\beta} x_1$, $ \nu $ is the effective frequency of collisions,
$\beta=m/(2kT)$, $m$ is the mass of a molecule of gas, $k $ is the
Boltzmann constant, $T $ is the temperature of gas,
constants $ \nu $, $ \gamma $ and $ \omega $ are expressed through
Prandtl number $ \Pr $, self-diffusion coefficient $D $ and kinematic
viscosity $\nu_*$  by following equalities \cite{eq}
$$
\nu=\dfrac{kT}{mD},\qquad \gamma=\dfrac{4}{5}\Big[1-\Big(1-
\dfrac{\omega}{2}\Big)\Pr\Big],\qquad
\omega=2\Big(1-\dfrac{D}{\nu_*}\Big).
$$

Besides, in the equation (2.1) $U_y (x)$ is the dimensionless mass
velocity of gas along an axis $y $,
$$
U_y(x)=\dfrac{1}{\pi^{3/2}}\int  \exp(-C'^2)C_y'
h(x,\mathbf{C'})d^3C',
$$
$Q_y(x)$ is the  сomponent of the vector of thermal
stream along an axis  $y$,
$$
{Q}_y(x)=\dfrac{1}{\pi^{3/2}}\int \exp(-C'^2)C_y'
\Big(C^2-\dfrac{5}{2}\Big)h(x,\mathbf{C'})d^3C',
$$
and $P_{xy}(x)$  is the component of viscous pressure tensor,
$$
P_{xy}(x)=\dfrac{1}{\pi^{3/2}}\int \exp(-C'^2)
C_x'C_y'h(x,\mathbf{C'})d^3C',
$$

Let's expand the function $h $ in two directions
$$
h(x,\mathbf{C})=C_yh_1(x,C_x)+\gamma
C_y\Big(C^2-\dfrac{5}{2}\Big)h_2(x,C_x).
\eqno{(2.2)}
$$

By means of (2.2) mass velocity is equal
$$
U_y(x)=\dfrac{1}{2\sqrt{\pi}}\int\limits_{-\infty}^{\infty}
e^{-\mu'^2}\Big[h_1(x,\mu')+\gamma\Big(\mu'^2-\dfrac{1}{2}\Big)
h_2(x,\mu')\Big]d\mu',
$$
$y$--component of thermal stream vector equals
$$
Q_y(x)=\dfrac{1}{2\sqrt{\pi}}\int\limits_{-\infty}^{\infty}e^{-\mu'^2}
\left\{\Big(\mu'^2-\dfrac{1}{2}\Big)h_1(x,\mu')+
\gamma\Big[\Big(\mu'^2-\dfrac{1}{2}\Big)^2+2\Big]\right\}d\mu',
$$
$xy$--component of viscous pressure tensor equals
$$
P_{xy}(x)=\dfrac{1}{2\sqrt{\pi}}\int\limits_{-\infty}^{\infty}
e^{-\mu'^2}\mu'\Big[h_1(x,\mu')+\gamma\Big(\mu'^2-\dfrac{1}{2}\Big)
h_2(x,\mu')\Big]d\mu'.
$$

By means of last three equalities and decomposition (2.2) we conclude,
that the equation (2.1) is equivalent to system from two equations
$$
\mu\dfrac{\partial h_1}{\partial x}+h_1(x,\mu)=
$$
$$=\dfrac{1}{\sqrt{\pi}}
\int\limits_{-\infty}^{\infty}e^{-\mu'^2}(1+\omega \mu\mu')
[h_1(x,\mu')+\gamma\Big(\mu'^2-\dfrac{1}{2}\Big)h_2(x,\mu')]d\mu'
$$
and
$$
\mu\dfrac{\partial h_2}{\partial x}+h_2(x,\mu)=$$$$=\dfrac{1}{2\sqrt{\pi}}
\int\limits_{-\infty}^{\infty}e^{-\mu'^2}
\Big\{\Big(\mu'^2-\dfrac{1}{2}\Big)h_1(x,\mu')+
\gamma\Big[\Big(\mu'^2-\dfrac{1}{2}\Big)^2+2\Big]h_2(x,\mu')\Big\}d\mu'.
$$

We introduce vector--column
$$
h(x,\mu)=\left(
           \begin{array}{c}
             h_1(x,\mu) \\
             h_2(x,\mu) \\
           \end{array}
         \right)
$$
and we transform previous system of equations in vector form
$$
\mu\dfrac{\partial h}{\partial x}+h(x,\mu)=\dfrac{1}{\sqrt{\pi}}
\int\limits_{-\infty}^{\infty}e^{-\mu'^2}K(\mu,\mu')h(x,\mu')d\mu'.
\eqno{(2.3)}
$$

In equation (2.3) $K(\mu,\mu')$ is the kernel of equation,
$$\extrarowheight=14pt
K(\mu,\mu')=\left(
              \begin{array}{cc}
1+\omega\mu\mu' & \gamma(\mu'^2-\dfrac{1}{2})(1+\omega\mu\mu') \\
                \dfrac{1}{2}(\mu'^2-\dfrac{1}{2}) & \dfrac{\gamma}{2}
                [(\mu'^2-\dfrac{1}{2})^2+2] \\
              \end{array}
            \right),
$$
with determinant
$$
\det K(\mu,\mu')=\gamma(1+\omega\mu\mu').
$$

The kernel of equation (2.3) we transform as sum
$$
K(\mu,\mu')=K(\mu')+\omega\mu\mu'\left(
                                   \begin{array}{cc}
                                     1 & \gamma(\mu'^2-\dfrac{1}{2}) \\
                                     0 & 0 \\
                                   \end{array}
                                 \right),
                                 $$
or, in the form
$$
K(\mu,\mu')=K(\mu')+\omega\mu\mu'\left(
                                   \begin{array}{cc}
                                     1 & 0 \\
                                     0 & 0 \\
                                   \end{array}
                         \right)K(\mu'),
$$
where
$$\extrarowheight=14pt
K(\mu')=\left(
              \begin{array}{cc}
1 & \gamma(\mu'^2-\dfrac{1}{2}) \\
                \dfrac{1}{2}(\mu'^2-\dfrac{1}{2}) & \dfrac{\gamma}{2}
                [(\mu'^2-\dfrac{1}{2})^2+2] \\
              \end{array}
            \right),
$$

$$
\det K(\mu')=\gamma.
$$

Now we will present the equation (2.3) in the form convenient for division
variables
$$
\mu\dfrac{\partial h}{\partial x}+h(x,\mu)=\dfrac{1}{\sqrt{\pi}}
\int\limits_{-\infty}^{\infty}e^{-\mu'^2}K(\mu')h(x,\mu')d\mu'+
$$
$$
+\omega\mu\dfrac{1}{\sqrt{\pi}}\int\limits_{-\infty}^{\infty}
e^{-\mu'^2}\mu'\left(
              \begin{array}{cc}
1 & 0 \\
0 & 0 \\
              \end{array}
            \right)K(\mu')h(x,\mu')d\mu'.
\eqno{(2.4)}
$$

\begin{center}
\bf  3. Division of variables. Dispersion matrix -- function
\end{center}

Following Euler, we search solutions of the equation (2.4) in the form
$$
h_\eta(x,\mu)=\exp(-\dfrac{x}{\eta})\Phi(\eta,\mu), \qquad
\eta\in \mathbb{C},
\eqno{(3.1)}
$$
where $\eta$ is the spectral parameter, generally speaking, complex
parameter.
Substituting (3.1) in the equation (2.4), we receive the characteristic
equation
$$
(\eta-\mu)\Phi(\eta,\mu)=\dfrac{1}{\sqrt{\pi}}\eta n(\eta)+
\omega\mu\left(
              \begin{array}{cc}
1 & 0 \\
0 & 0 \\
              \end{array}
            \right)m(\eta),
\eqno{(3.2)}
$$
in which
$$
n(\eta)=\int\limits_{-\infty}^{\infty}
e^{-\mu'^2}K(\mu')\Phi(\eta,\mu')d\mu',
\eqno{(3.3)}
$$
$$
m(\eta)=\int\limits_{-\infty}^{\infty}
e^{-\mu'^2}\mu'K(\mu')\Phi(\eta,\mu')d\mu',
$$

Multiplying the equation (3.2) at the left on the matrix $e^{-\mu^2}K(\mu)$
and integrating on all real axis, we receive that
$$
\left(
              \begin{array}{cc}
1 & 0 \\
0 & 0 \\
              \end{array}
            \right)m(\eta)=\left(
                             \begin{array}{c}
                               0 \\
                               0 \\
                             \end{array}
                           \right).
$$

Hence, the characteristic equation becomes simpler
$$
(\eta-\mu)\Phi(\eta,\mu)=\dfrac{1}{\sqrt{\pi}}\eta n(\eta).
\eqno{(3.4)}
$$

From the equation (3.4) by means of the condition of normalization (3.3)
we find eigen vectors of the characteristic equation correspond
to continuous spectrum $\sigma_c=(-\infty,+\infty)$
$$
\Phi(\eta,\mu)=\Big[\dfrac{1}{\sqrt{\pi}}\eta P\dfrac{1}{\eta-\mu}+
e^{\eta^2}K^{-1}(\eta)\Lambda(\eta)\delta(\eta-\mu)\Big]n(\eta).
\eqno{(3.5)}
$$

In (3.5) the symbol $P x^{-1}$ means principal value of integral
at integration of expression $x^{-1}$, $\delta(x)$ is the Dirac
delta--function, $\Lambda(z)$ is the dispersion matrix--fuction,
$$
\Lambda(z)=E+\dfrac{z}{\sqrt{\pi}}\int\limits_{-\infty}^{\infty}
\dfrac{e^{-\mu^2}K(\mu)}{\mu-z}d\mu,
$$
where $E$ is the unit matrix of the second order.

Let's present the dispersion matrix--function in the explicit form
$$
\Lambda(z)=$$
$$\extrarowheight=10pt
=\left(
             \begin{array}{cc}
               \lambda_0(z) & \gamma\lambda_0(z)(z^2-\dfrac{1}{2})+
               \dfrac{\gamma}{2} \\
               \dfrac{1}{2}\lambda_0(z)(z^2-\dfrac{1}{2})+
               \dfrac{1}{4} &\dfrac{\gamma}{2}[(z^2-\dfrac{1}{2})^2+2]
\lambda_0(z)+1+\dfrac{\gamma}{4}(z^2-\dfrac{9}{2}) \\
             \end{array}
           \right).
$$

Here $\lambda_0(z)$ is the dispersion plasma function,
$$
\lambda_0(z)=1+\dfrac{z}{\sqrt{\pi}}\int\limits_{-\infty}^{\infty}
\dfrac{e^{-\mu^2}d\mu}{\mu-z}.
$$

Let's present the dispersion matrix--function in the form, linear
concerning of the matrix $K(z)$
$$
\Lambda(z)=\lambda_0(z)K(z)+A(z),
$$
where
$$\extrarowheight=15pt
A(z)=\left(
       \begin{array}{cc}
         0 & \dfrac{\gamma}{2} \\
         \dfrac{1}{4} & 1+\dfrac{\gamma}{4}(z^2-\dfrac{9}{2}) \\
       \end{array}
     \right).
$$

Determinant of the dispersion matrix--functions we name
dispersion function. It is easy to see, that
$$
\lambda(z)\equiv\det\Lambda(z)=\gamma\lambda_0^2(z)+
\Big[1-\dfrac{\gamma}{4}\Big(z^2+\dfrac{7}{2}\Big)\Big]\lambda_0(z)-
\dfrac{\gamma}{8}.
$$

Let's expand dispersion function in asymptotic series in
neighbourhood of infinitely remote point
$$
\lambda(z)=\dfrac{1}{2}\Big(\dfrac{5\gamma}{4}-1\Big)\dfrac{1}{z^2}+
o\Big(\dfrac{1}{z^2}\Big), \qquad z\to \infty.
$$

This expansion means, that the discrete spectrum
of characteristic equation, consisting of zero of the dispersion
function, consists of one infinitely remote point
$z_i =\infty $ with order two. This spectrum is attached
to the continuous. This spectrum corresponds to
two solutions of the initial equations (2.4)
$$
h^{(1)}(x,\mu)=\left(\begin{array}{c}
                   1 \\
                   0 \\
                 \end{array}\right)
$$
and
$$
h^{(1)}(x,\mu)=\Big(x-\dfrac{2}{2-\omega}\mu\Big)\left(
                 \begin{array}{c}
                   1 \\
                   0 \\
                 \end{array}
               \right).
$$

\begin{center}
\bf  4. Boundary Kramers problem
\end{center}

On the condition of Kramers problem the given quantity is the gradient
of the mass velocity, given far from the wall
$$
g_v=\Big(\dfrac{du_y(x_1)}{dx_1}\Big)_{x_1=+\infty}.
\eqno{(4.1)}
$$

Let's consider, that for the dimensionless gradient is carried out
inequality $G_v\ll 1$ that allows to solve Kramers problem in
linear statement.

From the relation (4.1) we see, that far from the wall for mass
velocity is fair the asymptotic distribution
$$
u_y(x_1)=u_{sl}+g_vx_1, \qquad x_1\to +\infty.
$$

From here and from the equation (2.4) follows, that far from the wall function
$h(x,\mathbf{C})$ has following distribution
$$
h_{as}(x,\mathbf{C})=2C_yU_{sl}+2C_y(x-\dfrac{2}{2-\omega}C_x)G_v,
\qquad x\to +\infty.
\eqno{(4.2)}
$$

It is the linear combination of two partial solution of the initial
equation (2.1)
$$
h_1(x,\mathbf{C})=1
$$
and
$$
h_2(x,\mathbf{C})=x-\dfrac{2}{2-\omega}C_x.
$$

The distribution (4.2) is the Chapman---Enskog distribution.

By means of (2.2) we will present distribution (4.2) in the vector form
$$
h_{as}(x,\mu)=\Big[2U_{sl}+\Big(x-\dfrac{2}{2-\omega}\mu\Big)G_v\Big]
\left(\begin{array}{c}1 \\0 \\\end{array}\right).
\eqno{(4.3)}
$$

Decomposition (4.3) represents the linear combination of two
discrete (partial) solutions of the equation (2.4),
correspond to the spectrum attached to continuous spectrum.

Let's formulate boundary conditions in the Kramers problem under the condition
of diffusion reflexion of molecules from the wall
$$
h(0,\mu)=\left(\begin{array}{c}0 \\0 \\\end{array}\right), \qquad
\mu>0,
\eqno{(4.4)}
$$
$$
h(x,\mu)=h_{as}(x,\mu)+o(1),\qquad x\to +\infty.
\eqno{(4.5)}
$$

So, boundary Kramers problem consists in finding of such
solutions of the equation (2.4) which satisfies boundary
conditions (4.4) and (4.5).

The solution of the problem (2.4), (4.4) and (4.5) we search in the form
of the sum of the partial solutions of the attached spectrum and
integral on the continuous spectrum of continuous eigen
solutions
$$
h(x,\mu)=h_{as}(x,\mu)+\int\limits_{0}^{\infty}\exp(-\dfrac{x}{\eta})
\Phi(\eta,\mu)d\eta.
\eqno{(4.6)}
$$

Let's present the solution (4.6) in the explicit form
$$
h(x,\mu)=h_{as}(x,\mu)+\dfrac{1}{\sqrt{\pi}}\int\limits_{0}^{\infty}
\exp(-\dfrac{x}{\eta})\dfrac{\eta n(\eta)d\eta}{\eta-\mu}+
$$
$$
+\exp(\mu^2-\dfrac{x}{\mu})K^{-1}(\mu)\Lambda(\mu)n(\mu).
\eqno{(4.7)}
$$

Unknown members in expansion  (4.6) or (4.7) are dimensionless
velocity  of sliding $U_{sl}$ (coefficient of the attached spectrum) and
vector--function $n(\eta)$ (coefficient of the continuous spectrum).

Expansion (4.7) automatically satisfies to the condition (4.5).
Substituting (4.7) in the condition (4.4), we receive the one-side
vector singular integral equation with Cauchy kernel
$$
h_{as}(0,\mu)+\dfrac{1}{\sqrt{\pi}}\int\limits_{0}^{\infty}
\dfrac{\eta n(\eta)d\eta}{\eta-\mu}+e^{\mu^2}K^{-1}(\mu)\Lambda(\mu)
n(\mu)={\bf 0},\quad \mu>0.
\eqno{(4.8)}
$$

We introduce the matrix $P(z)=K^{-1}(z)\Lambda(z)$. It is easy find that
$$
P(z)=\lambda_0(z)E+B(z),
$$
where
$$ \extrarowheight=10pt
B(z)=\left(
       \begin{array}{cc}
         -\dfrac{1}{4}(z^2-\dfrac{1}{2}) & (\gamma-1)z^2+\dfrac{1}{2} \\
         \dfrac{1}{4\gamma} & \dfrac{1}{\gamma}-1 \\
       \end{array}
     \right),\qquad \det B(z)=-\dfrac{1}{8}.
$$

For the matrix $P(z)$ formulas Sokhotsky are carried out
$$
P^+(\mu)-P^-(\mu)=2\sqrt{\pi}i\mu e^{-\mu^2}, \qquad
-\infty<\mu<+\infty,
$$
$$
\dfrac{P^+(\mu)+P^-(\mu)}{2}=P(\mu), \qquad
P(\mu)=\lambda_0(\mu)E+B(\mu).
$$

The singular integral equation (4.8) we transform to the vector
boundary condition
$$
P^+(\mu)[h_{as}(0,\mu)+N^+(\mu)]=P^-(\mu)[h_{as}(0,\mu)+N^-(\mu)],
\quad \mu>0.
\eqno{(4.9)}
$$

In (4.9) we have entered the new auxiliary vector--function

$$
N(z)=\dfrac{1}{\sqrt{\pi}}\int\limits_{0}^{\infty}
\dfrac{\eta n(\eta)d\eta}{\eta-z},
\eqno{(4.10)}
$$
for which Sokhotsky formulas are carried out
$$
N^+(\mu)-N^-(\mu)=2\sqrt{\pi}i \mu n(\mu), \qquad \mu>0,
\eqno{(4.10a)}
$$
$$
\dfrac{1}{2}[N^+(\mu)+N^-(\mu)]=N(\mu),\qquad \mu>0.
$$

Boundary condition (4.9) we will present in the form of the non-uniform vector
boundaty value Riemann---Hilbert problem with matrix coefficient
$$
N^+(\mu)=G(\mu)N^-(\mu)+[G(\mu)-E]h_{as}(0,\mu), \qquad \mu>0.
\eqno{(4.11)}
$$

Coefficient of the problem (4.11) is the matrix--function
$$
G(\mu)=[P^+(\mu)]^{-1}P^-(\mu)=[\Lambda^+(\mu)]\Lambda^-(\mu).
$$

\begin{center}
  \bf 5. Homogeneous vector boundary value Riemann---Hilbert problem
\end{center}

For solution of the problem (4.11) at first we will solve the corresponding
homogeneous vector boundary value problem
$$
X^+(\mu)=G(\mu)X^-(\mu), \qquad \mu>0.
\eqno{(5.1)}
$$
where the unknown matrix--function $X(z)$ is analytic in complex
planes with a cut along the real positive half-axis.

For solution of this problem reduction to diagonal form the matrix
$P(z)$ is required. For this purpose reduction to diagonal form
the matrix $B(z)$ is required.

Eigen numbers (functions) of matrix $B(z)$ are equal
$$
\mu_{1,2}(z)=-\dfrac{1}{8}\Big(z^2+\dfrac{7}{2}-\dfrac{4}{\gamma}\Big)
\pm \dfrac{1}{8}r(z).
$$

Here
$$
r(z)=\sqrt{q(z)},\qquad
q(z)=\Big(z^2+\dfrac{7}{2}-\dfrac{4}{\gamma}\Big),
$$
where low index 1 corresponds to sign plus, and index 2 corresponds to
sign minus.

The matrix transforming the matrix $B(z)$ to  diagonal form, has
the following form
$$
S(z)=\left(
       \begin{array}{cc}
         \mu_1(z)-\dfrac{1}{\gamma}+1 & \mu_2(z)-\dfrac{1}{\gamma}+1 \\
         \dfrac{1}{4\gamma} & \dfrac{1}{4\gamma} \\
       \end{array}
     \right), \qquad \det S(z)=\dfrac{r(z)}{16\gamma}.
$$

More low the return matrix is required to us also
$$\extrarowheight=14pt
S^{-1}(z)=\dfrac{4}{r(z)}\left(
                           \begin{array}{cc}
1& \dfrac{\gamma}{2}\Big[r(z)+\Big(z^2-\dfrac{9}{2}+
\dfrac{4}{\gamma}\Big)\Big] \\
-1&  \dfrac{\gamma}{2}\Big[r(z)-\Big(z^2-\dfrac{9}{2}+
\dfrac{4}{\gamma}\Big)\Big]  \\
                           \end{array}
                         \right).
$$

For the solution of the problem (5.1) we search in the form
$$
X(z)=S(z)U(z)S^{-1}(z),
\eqno{(5.2)}
$$
where $U(z)$ is the new unknown diagonal matrix,
$$
U(z)=\left(
       \begin{array}{cc}
         U_1(z) & 0 \\
         0 & U_2(z) \\
       \end{array}
     \right)=\diag\{U_1(z), U_2(z)\}.
$$

Matrix $S(z)$ and return to it contain the radical $r(z)$,
representing square root from the polynom of the fourth
degree $q(z)$. Hence, these matrixes have four branching points.
These points are polynom zero $q(z)$:
$$
a(\gamma),\qquad \bar a(\gamma),\qquad-a(\gamma),\qquad-\bar a(\gamma),
$$
where
$a(\gamma) $ is the zero laying in the first quarter,
$$
a(\gamma)=\sqrt{\dfrac{4}{\gamma}-\dfrac{7}{2}+i\sqrt{8}}.
$$

Additional cuts we will spend as follows. We will connect
points of branching with infinitely remote point following beams
$$
\Gamma_1(\gamma)=[a(\gamma),+\infty+i\sqrt[4]{8}],\qquad
\Gamma_2(\gamma)=[-\bar a(\gamma),-\infty+i\sqrt[4]{8}],
$$
$$\Gamma_3(\gamma)=[\bar a(\gamma),+\infty-i\sqrt[4]{8}],\qquad
\Gamma_4(\gamma)=[-a(\gamma),-\infty-i\sqrt[4]{8}].
$$

Let's unite these cuts, having entered the designation
$$
\Gamma(\gamma)=\bigcup\limits_{j=1}^{j=4}\Gamma_j(\gamma).
$$

Matrixes $S(z)$ and $S^{-1}(z)$ are analytical in all complex
planes with the cut lengthways $\Gamma(\gamma)$. It means, that
we search unequivocal analytical matrix  $X(z)$ in domain
$\mathbb{C}\setminus \Big(\Gamma(\gamma)\bigcup \overline{\mathbb
R}\Big)$.

Substituting (5.2) in (5.1), we receive the following matrix
boundary value problem
$$
\Omega^+(\mu)U^+(\mu)=\Omega^-(\mu)U^-(\mu), \qquad \mu>0,
\eqno{(5.3)}
$$
where the matrix $\Omega(z)$  is entered by following equality
$$
\Omega(z)=S^{-1}(z)P(z)S(z)=\lambda_0(z)E+S^{-1}(z)B(z)S(z)=
$$
$$
=\lambda_0(z)E+\diag\{\mu_1(z), \mu_2(z)\}=\diag\{\Omega_1(z),
\Omega_2(z)\},
$$
where
$$
\Omega_j(z)=\lambda_0(z)+\mu_1(z)=\lambda_0(z)-\dfrac{1}{8}
\Big(z^2+\dfrac{7}{2}-\dfrac{4}{\gamma}\Big)\pm\dfrac{1}{8}
\sqrt{q(z)},\quad j=1,2,
$$
and $j=1$ corresponds to the sign plus, and $j=2$  corresponds to the
sign minus.

Considering, that the structure of matrix  $X(z)$ contains radicals
$\pm r(z)$, changing the sign at transition  on opposite  through
additional cuts $\Gamma_j(\gamma) \; (j=1,2,3,4)$, for
unambiguity of a matrix $X(z)$ should be demanded, that on
coast of additional cuts boundary values of the matrix
$X(z)$ from above and from below were equal
$$
X^+(\tau)=X^-(\tau), \qquad \tau\in \Gamma(\gamma),
\eqno{(5.4)}
$$
or, that all the same,
$$
U^+(\tau)T(\tau)=T(\tau)U^-(\tau),\qquad \tau\in \Gamma(\gamma),
\eqno{(5.5)}
$$
where the matrix $T(\tau)$ is defined on the additional cut and
has the following form
$$
T(\tau)=\Big[S^+(\tau)\Big]^{-1}S^-(\tau),\qquad \tau\in
\Gamma(\gamma).
$$

Rectilinear calculations show, that a matrix $T(\tau)$
is constant
$$
T(\tau)=\left(
          \begin{array}{cc}
            0 & 1 \\
            1 & 0 \\
          \end{array}
        \right).
$$

The matrix boundary value problem (5.3) is equivalent to two scalar
boundary value problems on the basic cut
$$
U^+_j(\mu)=\dfrac{\Omega_j^-(\mu)}{\Omega_j^+(\mu)}U_j^-(\mu),
\qquad j=1,2, \qquad \mu>0.
$$

Noticing, that
$$
\overline{\Omega_j^-(\mu)}=\Omega_j^+(\mu), \qquad
-\infty<\mu<+\infty,
$$
and сonsidering augmentation of arguments
$\theta_j(\mu)=\arg \Omega^+(\mu)$ on the semi\-axis $[0,+ \infty]$,
we will rewrite these problems in the following form
$$
U_1^+(\mu)=\exp(-2i \theta_1(\mu))U_1^-(\mu),\qquad \mu>0,
\eqno{(5.6)}
$$
and
$$
U_2^+(\mu)=\exp(-2i [\theta_2(\mu)-\pi])U_2^-(\mu),\qquad \mu>0,
\eqno{(5.7)}
$$

Let's notice, that problems (5.4) (or (5.5)) are not reduced to scalar
problems, and
are essentially vector boundary value problems rather in respect of
vector $U(z)=\{U_1(z),U_2(z)\} $.
By means of the found matrix $T $
let's rewrite these problems in the form of two vector boundary
value problems
$$
U_1^+(\tau)=U_2^-(\tau),\qquad \tau\in \Gamma(\gamma),
\eqno{(5.8)}
$$
$$
U_1^-(\tau)=U_2^+(\tau),\qquad \tau\in \Gamma(\gamma).
\eqno{(5.9)}
$$

Now the basic difficulty consists in search of such solution
$$
U(z)=\{U_1(z),U_2(z)\},
$$
which would satisfy simultaneously
to four boundary value problems (5.6) -- (5.9).

Let's proceed  as follows. We will multiply and will divide
against each other boundary value problems (5.6) and (5.7).
Then we find the logarithm of the received problems, and the second of
them we will divide term by term on
radical $r(z)$. We receive, that
$$
\ln[U_1(\mu)U_2(\mu)]^+-\ln[U_1(\mu)U_2(\mu)]^-=-2i[\theta_1(\mu)+
\theta_2(\mu)-\pi], \qquad \mu>0,
$$
and
$$
\dfrac{1}{r(\mu)}\ln\Bigg[\dfrac{U_1(\mu)}{U_2(\mu)}\Bigg]^+-
\dfrac{1}{r(\mu)}\ln\Bigg[\dfrac{U_1(\mu)}{U_2(\mu)}\Bigg]^-
=-2i\dfrac{\theta_1(\mu)-\theta_2(\mu)+\pi}{r(\mu)}, \quad
\mu>0.
$$

These problems as problems "on jump"\,, have the following solutions
$$
\ln\Big[U_1(z)U_2(z)\Big]=-\dfrac{1}{\pi}\int\limits_{0}^{\infty}
\dfrac{\theta_1(\mu)+\theta_2(\mu)-\pi}{\mu-z}d\mu,
$$
and
$$
\dfrac{1}{r(z)}\ln\Bigg[\dfrac{U_1(z)}{U_2(z)}\Bigg]=
-\dfrac{1}{\pi}\int\limits_{0}^{\infty}
\dfrac{\theta_1(\mu)-\theta_2(\mu)+\pi}{r(\mu)(\mu-z)}d\mu.
$$

From two last equalities we receive
$$
U_1(z)U_2(z)=\exp(-2A(z))
$$
and
$$
\dfrac{U_1(z)}{U_2(z)}=\exp(-2r(z)B(z)),
$$
where
$$
A(z)=\dfrac{1}{2\pi}\int\limits_{0}^{\infty}
\dfrac{\theta_1(\mu)+\theta_2(\mu)-\pi}{\mu-z}d\mu,
$$
and
$$
B(z)=\dfrac{1}{2\pi}\int\limits_{0}^{\infty}
\dfrac{\theta_1(\mu)-\theta_2(\mu)+\pi}{r(\mu)(\mu-z)}d\mu.
$$

Hence, having designated the received solution through
$$
U^\circ(z)=\{U_1^\circ(z), U_2^\circ(z)\},
$$
we will write
$$
U_1^\circ(z)=\exp(-A(z)-r(z)B(z)),
$$
and
$$
U_2^\circ(z)=\exp(-A(z)+r(z)B(z)).
$$

It is easy to check up, that these functions are the solution at once all
four boundary value problems (5.9) -- (5.9). However, the received
solution has one basic lack, which is essential
singularity in infinitely remote point.

For its elimination we search functions $U_j(z) (j=1,2) $ in the form
$$
U_1(z)=U_1^\circ(z)\cdot \varphi(z)
$$
and
$$
U_2(z)=U_2^\circ(z)\cdot\dfrac{1}{\varphi(z)},
$$
where function $\varphi(z) $ is analitycal in the complex plane out of
additional cuts $\Gamma(\gamma)$ (with the essential
singularity in infinitely remote point).
Thus boundary value conditions (5.6) and (5.7) are carried out
automatically, and boundary value conditions (5.8) and (5.9) are
carried out then and only when on additional cuts is carried out
condition
$$
\varphi^+(\tau)=\dfrac{1}{\varphi^-(\tau)},\qquad \tau\in
\Gamma(\gamma).
$$

As the solution of this nonlinear boundary value problem we take
function
$$
\varphi(z)=\exp(r(z)R(z)),
$$
where
$$
R(z)=\int\limits_{0}^{\mu_0}\dfrac{d\tau}{r(\tau)(\tau-z)}.
$$

Here the point $\mu_0\in (0,+\infty) $ is more low defined unequivocally.

Without the proof we will inform (see, for example, \cite{23}),
that the point $ \mu_0$ is the unique solution of the special case
 Jacobi problem  of inverse for elliptic integrals:
$$
\dfrac{1}{2\pi}\int\limits_{0}^{\infty}\dfrac{\theta_1(\mu)-
\theta_2(\mu)+\pi}{r(\mu)}d\mu=\int\limits_{0}^{\mu_0}
\dfrac{d\tau}{r(\tau)}.
\eqno{(5.10)}
$$

It is possible to check up, that functions
$$
U_1(z)=\exp\Big[-A(z)-r(z)\Big(B(z)-R(z)\Big)\Big]
\eqno{(5.11)}
$$
and
$$
U_2(z)=\exp\Big[-A(z)+r(z)\Big(B(z)-R(z)\Big)\Big]
\eqno{(5.12)}
$$
are solutions at once all four boundary value problems (5.6) -- (5.9)
and also have no essential singularity in infinitely removed
point at performance of the condition (5.10).

It is possible to show, that function $U_1(z) $ has a simple pole in
the origin of coordinates and simple zero in the point $ \mu_0$, and function
$U_2(z) $ is limited in the origin of coordinates and does not
disappear, and has simple pole in the point $\mu_0$.

So, factor--matrix $X (z) $ is constructed and has following elements
$$
X_{11}(z)=\dfrac{U_1(z)+U_2(z)}{2}-\dfrac{z^2-\dfrac{9}{2}+\dfrac{4}{\gamma}}
{r(z)}\dfrac{U_1(z)-U_2(z)}{2},
$$
$$
X_{12}(z)=\dfrac{2}{r(z)}\Big[2z^2(\gamma-1)+1\Big](U_1(z)-U_2(z)),
$$
$$
X_{21}(z)=\dfrac{U_1(z)-U_2(z)}{\gamma r(z)},
$$
$$
X_{22}(z)=\dfrac{U_1(z)+U_2(z)}{2}+\dfrac{z^2-\dfrac{9}{2}+\dfrac{4}{\gamma}}
{r(z)}\dfrac{U_1(z)-U_2(z)}{2}.
$$

On construction, the matrix--function $X(z)$ satisfies to the condition
(5.1), is analytical everywhere in $ \mathbb{C}\setminus [0,+\infty]$ with
possible special points in polynom zero $q(z)$. But thanking
coincidence of functions $U_1(z)$ and $U_2(z)$ in polynom zero $q(z)$
matrix $X(z)$ is analytical and in these points. The matrix deter\-mi\-nant
$X(z)$ does not degenerate everywhere in $ \mathbb{C}\setminus [0, +\infty]$
also has a pole of the first order in the origin of coordinates.

So, homogeneous boundary value Riemann---Hilbert  problem  (5.1)
is comp\-le\-tely solved.

\begin{center}
  \bf 6. Inhomogeneous vector boundary value Riemann---Hilbert problem
  and its conditions of resolvability
\end{center}

By means of the factorization problem (5.1) we will reduce the problem (4.9) to
vector boundary value Riemann---Hilbert problem "on zero jump"
$$
\Big[X^+(\mu)\Big]^{-1}\Big[N^+(\mu)+h_{as}(0,\mu)\Big]=$$$$=
\Big[X^-(\mu)\Big]^{-1}\Big[N^-(\mu)+h_{as}(0,\mu)\Big], \quad
\mu>0.
\eqno{(6.1)}
$$

Here
$$
X^{-1}(z)=S(z)U^{-1}(z)S^{-1}(z).
$$

Considering behaviour of matrixes and vectros entering into the
boundary condition (6.1) in the complex plane, we will write its
common solution
$$
X(z)=-h_{as}(0,z)+X(z)\Phi(z),
\eqno{(6.2)}
$$
where
$$
\Phi(z)=\left(
          \begin{array}{c}
            \Phi_1(z) \\
            \Phi_2(z) \\
          \end{array}
        \right)=\left(
                  \begin{array}{c}
                    \alpha_1z+\alpha_0+\dfrac{\alpha_{-1}}{z-\mu_0} \\
                    \beta_1z+\beta_0+\dfrac{\beta_{-1}}{z-\mu_0} \\
                  \end{array}
                \right),
\eqno{(6.3)}
$$
and in expression (6.3) all coefficients $ \alpha_j, \beta_j \;
(j =-1,0,1) $ are arbitrary constants.

The received solution (6.2) has following singularities: the pole
the first order in the origin of coordinates $z=0$, the pole of the second order
in the point $z =\mu_0$ and the pole of the
first order in the point $z =\infty $.

That the vector $N(z)$, defined by equality (6.2), was possible
to accept as auxiliary function $N(z)$, entered
above by equality (4.10), we will eliminate the specified
singularities for the account
choice of free parametres of the solution (6.2) and the unknown
coefficient  of  attached spectrum $U_{sl}$.

Let's rewrite the solution (6.2) in the explicit form
$$
N(z)=-\Big(U_{sl}-\dfrac{2G_v}{2-\omega}z\Big)\left(
                                               \begin{array}{c}
                                                 1 \\
                                                 0 \\
                                               \end{array}
                                             \right)+
S(z)U(z)S^{-1}(z)\Phi(z).
\eqno{(6.4)}
$$

Let's find asymptotic of matrix $X(z)$ in the vicinity infinitely
remote point
$$ \extrarowheight=14pt
X(z)=\left(
       \begin{array}{cc}
         U_2(z) & 4(\gamma-1)[U_1(z)-U_2(z)] \\
         0 & U_1(z) \\
       \end{array}
     \right)+o\Big(\dfrac{1}{z}\Big),\qquad z\to \infty.
\eqno{(6.5)}
$$

Let's expand functions $A(z), B(z)$ and $R(z)$ into Laurent series in
vicinities of infinitely remote point
$$
A(z)=\dfrac{A_{-1}}{z}+\dfrac{A_{-2}}{z^2}+\cdots, \qquad z\to
\infty,
$$
$$
B(z)=\dfrac{B_{-1}}{z}+\dfrac{B_{-2}}{z^2}+\cdots, \qquad z\to
\infty,
$$
$$
R(z)=\dfrac{R_{-1}}{z}+\dfrac{R_{-2}}{z^2}+\cdots, \qquad z\to
\infty.
$$

Here
$$
A_{-k}=-\dfrac{1}{2\pi}\int\limits_{0}^{\infty}\tau^{k-1}
[\theta_1(\tau)+\theta_2(\tau)-\pi]d\tau,\qquad k=1,2,\cdots,
$$
$$
B_{-k}=-\dfrac{1}{2\pi}\int\limits_{0}^{\infty}\tau^{k-1}
\dfrac{\theta_1(\tau)-\theta_2(\tau)+\pi}{r(\tau)}d\tau,
\qquad k=1,2,\cdots,
$$
$$
R_{-k}=-\int\limits_{0}^{\mu_0}\dfrac{\tau^{k-1}}{r(\tau)}d\tau,
\qquad k=1,2,\cdots.
$$

Now, considering equalities (5.11) and (5.12), it is easy to expand
into Laurent series functions $U_1(z)$ and $U_2(z)$
$$
U_1(z)=p_0\Big(1+\dfrac{p_{-1}}{z}+\cdots\Big), \qquad z\to
\infty,
$$
$$
U_2(z)=q_0\Big(1+\dfrac{q_{-1}}{z}+\cdots\Big), \qquad z\to
\infty.
$$

Here
$$
p_0=\exp[-(B_{-2}-R_{-2})], \qquad q_0=\dfrac{1}{p_0},
$$
$$
p_{-1}=-A_{-1}-(B_{-3}-R_{-3}),
$$
$$
q_{-1}=-A_{-1}+(B_{-3}-R_{-3}).
$$

In terms of Laurent coefficients of functions $B(z)$ and $R(z)$
the problem of Jacobi inverse (5.10) can be presented in the form
simple equality
$$
B_{-1}=R_{-1}.
\eqno{(6.6)}
$$

According to asymptotic equality (6.5), we have:
$$
X(z)=X_0+X_{-1}\dfrac{1}{z}+o\Big(\dfrac{1}{z}\Big),\qquad z\to
\infty.
\eqno{(6.7)}
$$

Here
$$ \extrarowheight=14pt
X_0=\left(
      \begin{array}{cc}
        q_0 & 4(\gamma-1)(p_0-q_0) \\
        0 & p_0 \\
      \end{array}
    \right)
$$

and

$$ \extrarowheight=14pt
X_{-1}=\left(
      \begin{array}{cc}
        q_0q_{-1} & 4(\gamma-1)(p_0p_{-1}-q_0q_{-1}) \\
        0 & p_0p_{-1} \\
      \end{array}
    \right).
$$

On the basis of (6.7) we will write out decomposition top and bottom
elements of common solution  in the vicinity infinitely removed
point
$$
N_1(z)=-2\Big(U_{sl}-\dfrac{2G_v}{2-\omega}z\Big)+X_{11}(z)
\Big(\alpha_1z+\alpha_0+\dfrac{\alpha_{-1}}{z-\mu_0}\Big)+$$$$+
X_{12}(z)\Big(\beta_1z+\beta_0+\dfrac{\beta_{-1}}{z-\mu_0}\Big)
+O(\dfrac{1}{z}),
\eqno{(6.8)}
$$
$$
N_2(z)=
X_{22}(z)\Big(\beta_1z+\beta_0+\dfrac{\beta_{-1}}{z-\mu_0}\Big)+
O(\dfrac{1}{z}).
\eqno{(6.9)}
$$

From decomposition (6.8) and (6.9) taking into account the
previous equalities it is visible, that
$$
\beta_0=\beta_1=0,
$$
and
$$
\alpha_1=-\dfrac{4G_v}{2-\omega}\cdot\dfrac{1}{q_0},
\eqno{(6.10)}
$$
$$
U_{sl}=-\dfrac{2G_v}{2-\omega}\cdot\dfrac{q_{-1}}{q_0}+
\dfrac{1}{2}q_0\alpha_0.
\eqno{(6.11)}
$$

The pole of the second order in the point $ \mu_0$ has the bottom element
of the common solution. Hence, for elimination of this pole
it is necessary and sufficient to demand that it was carried out
equality:
$$
S^{-1}_{21}(z)\Phi_1(z)+S^{-1}_{22}(z)\Phi_2(z)=O(z-\mu_0),\qquad
z\to \mu_0.
$$

Let's write down this condition in the explicit form
$$
-\Big(\alpha_1z+\alpha_0+\dfrac{\alpha_{-1}}{z-\mu_0}\Big)+
\dfrac{\gamma}{2}\Big[r(z)-\Big(z^2-\dfrac{9}{2}+
\dfrac{4}{\gamma}\Big)\Big]\dfrac{\beta_{-1}}{z-\mu_0}=O(z-\mu_0),
z\to \mu_0.
$$
From here we find, that
$$
\alpha_1=\alpha(\mu_0)\beta_{-1},
$$
and
$$
\alpha_1\mu_0+\alpha_{0}=\alpha'(\mu_0)\beta_{-1}.
\eqno{(6.12)}
$$
Here
$$
\alpha(z)=\dfrac{\gamma}{2}\Big[r(z)-\Big(z^2-\dfrac{9}{2}+
\dfrac{4}{\gamma}\Big)\Big],
$$
$$
\alpha'(z)=\dfrac{\gamma}{2}z\Big[\dfrac{3z^2+2(7-{8}/{\gamma})}
{2r(z)}-2\Big].
$$

For pole elimination in zero it is necessary and sufficient
to demand, that equality was carried out
$$
S^{-1}_{11}(z)\Phi_1(z)+S^{-1}_{12}(z)\Phi_2(z)=O(z),\qquad
z\to 0.
$$

From this relation we find that
$$
\beta_{-1}=\dfrac{\mu_0\alpha_0}{\alpha(\mu_0)+\delta},
\eqno{(6.13)}
$$
where
$$
\delta=\dfrac{\gamma}{2}\Big[r(0)-\dfrac{9}{2}+\dfrac{4}{\gamma}\Big],
$$
$$
r(0)=\sqrt{\dfrac{81}{9}+\dfrac{16}{\gamma^2}-\dfrac{28}{\gamma}}=
\sqrt{\Big(\dfrac{7}{2}-\dfrac{4}{\gamma}\Big)^2+8}=
\sqrt{\Big(\dfrac{9}{2}-\dfrac{4}{\gamma}\Big)^2+\dfrac{8}{\gamma}}.
$$

Substituting (6.13) in (6.12), we find
$$
\alpha_0=-\dfrac{\alpha_1\mu_0(\alpha(\mu_0)+\delta)}
{\alpha(\mu_0)+\delta-\mu_0\alpha'(\mu_0)}.
\eqno{(6.14)}
$$

Now on the basis of (6.11) by means of (6.10) and (6.14) we found
required velocity of sliding (the coefficient correspond to the attached
spectrum)
$$
U_{sl}=-\dfrac{2G_v}{2-\omega}\Big[q_{-1}-
\dfrac{\mu_0(\alpha(\mu_0)+\delta)}{\alpha(\mu_0)+\delta-\mu_0
\alpha'(\mu_0)}\Big].
\eqno{(6.15)}
$$

The coefficient  $n(\mu)$, correspond to the continuous spectrum,
we find from equality (4.10а):
$$
n(\mu)=\dfrac{1}{2\sqrt{\pi}i\mu}[{N^+(\mu)-N^-(\mu)}]=$$$$
\qquad \quad\quad=
\dfrac{1}{2\sqrt{\pi}i\mu}[X^+(\mu)-X^-(\mu)]\Phi(\mu).
\eqno{(6.16)}
$$

So, all coefficients of decomposition (4.6) are found.
Coefficient of continuous spectrum is given by equality (6.16), and
coefficient of the spectrum attached to continuous, it is given by equality
(6.15).

\begin{center}
  \bf 7. Conclusion
\end{center}

In the present work the analytical solution of the Kramers problem
about isothermal sliding for the Holway---Shakhov equation
is constructed.

\end{document}